\documentclass[conference]{IEEEtran}
\IEEEoverridecommandlockouts
\usepackage{cite}
\usepackage{amsmath,amssymb,amsfonts}
\usepackage{algorithmic}
\usepackage{graphicx}
\usepackage{textcomp}
\usepackage{xcolor}
\usepackage{ bbold }
\usepackage{subfigure}
\usepackage[linesnumbered,ruled,vlined]{algorithm2e}
\usepackage{multirow}
\graphicspath{ {images/} }
\def\BibTeX{{\rm B\kern-.05em{\sc i\kern-.025em b}\kern-.08em
    T\kern-.1667em\lower.7ex\hbox{E}\kern-.125emX}}

\makeatother
\begin{document}

\title{Computer-Aided Clinical Skin Disease Diagnosis Using CNN and Object Detection Models}

\author{
\IEEEauthorblockN{
Xin He\IEEEauthorrefmark{1}, 
Shihao Wang\IEEEauthorrefmark{1}, 
Shaohuai Shi\IEEEauthorrefmark{1}, 
Zhenheng Tang\IEEEauthorrefmark{1}, \\
Yuxin Wang\IEEEauthorrefmark{1}, 
Zhihao Zhao\IEEEauthorrefmark{1}, 
Jing Dai\IEEEauthorrefmark{1}, 
Ronghao Ni\IEEEauthorrefmark{1}, \\
Xiaofeng Zhang\IEEEauthorrefmark{2},
Xiaoming Liu\IEEEauthorrefmark{3},
Zhili Wu\IEEEauthorrefmark{4}, 
Wu Yu\IEEEauthorrefmark{4},
Xiaowen Chu\thanks{\IEEEauthorrefmark{5}Corresponding author: chxw@comp.hkbu.edu.hk}\IEEEauthorrefmark{1}\IEEEauthorrefmark{5}}

\IEEEauthorblockA{
\IEEEauthorrefmark{1}Department of Computer Science, Hong Kong Baptist University\\
\IEEEauthorrefmark{2}Harbin Institute of Technology, China\\
\IEEEauthorrefmark{3}The University of Hong Kong - Shenzhen Hospital, China\\
\IEEEauthorrefmark{4}Shenzhen Yiyuan Intelligent Technology Co., Ltd., China
}



}



\maketitle


\begin{abstract}
Skin disease is one of the most common types of human diseases, which may happen to everyone regardless of age, gender or race. Due to the high visual diversity, human diagnosis highly relies on personal experience; and there is a serious shortage of experienced dermatologists in many countries. To alleviate this problem, computer-aided diagnosis with state-of-the-art (SOTA) machine learning techniques would be a promising solution. In this paper, we aim at understanding the performance of convolutional neural network (CNN) based approaches. We first build two versions of  skin disease datasets from Internet images: (a) Skin-10, which contains 10 common classes of skin disease with a total of 10,218 images; (b) Skin-100, which is a larger dataset that consists of 19,807 images of 100 skin disease classes. Based on these datasets, we benchmark several SOTA CNN models and show that the accuracy of skin-100 is much lower than the accuracy of skin-10. We then implement an ensemble method based on several CNN models and achieve the best accuracy of 79.01\% for Skin-10 and 53.54\% for Skin-100. We also present an object detection based approach by introducing bounding boxes into the Skin-10 dataset. Our results show that object detection can help improve the accuracy of some skin disease classes. 

\end{abstract}


\section{Introduction}

It is generally known that there are many types of skin diseases, ranging from itching caused by mosquito bites to skin cancer. Traditionally, the diagnosis of skin diseases is based on the comprehensive consideration of the size, shape, color and other visual features of the lesion area. Recently, deep learning methods have been applied to many medical tasks \cite{rajpurkar2017mura,moreira2012inbreast} and obtained remarkable achievements. Using computers to help diagnose skin diseases would be a promising direction. At present, there are many studies \cite{mendoncca2013ph,tschandl2018ham10000} on dermoscopic images and they have achieved promising results. Although there are some works \cite{razeghi2012skin,sun2016benchmark} on clinical skin disease images, their datasets are small. To fill this gap, we first build a clinical skin disease dataset, namely Skin-10, which has 10,218 images of 10 common skin diseases, and we manually marked the skin lesions for each image. Besides, we expand Skin-10 to Skin-100, which covers 100 classes of skin diseases and contains 19,807 clinical images. To the best of our knowledge, the scale of Skin-100 is larger than all existing clinical skin disease datasets. What's more, Skin-10 is the first clinical skin disease dataset that provides bounding boxes. By using the bounding boxes, we could extract more discriminative features and thereby improve the classification performance. 



\begin{table*}[!hbt]
    \centering
    \caption{The summary of existing skin disease datasets. In the first row (Type), D and C indicates dermoscopic and clinical, respectively. The fourth row (Annotation?) represents whether the dataset provides bounding boxes or segmentation. }\label{table:dataset_summary}
    \begin{tabular}{|c|c|c|c|c|c|c|c|c|c|}
        \hline
        Dataset & 
        PH2\cite{mendoncca2013ph}& 
        ISIC & 
        Ham10000\cite{tschandl2018ham10000}   &
        \cite{data1998} & \cite{razeghi2012skin}$-1^{st}$ / $-2^{nd}$ & \cite{razeghi2013interactive} & \cite{sun2016benchmark}$-1^{st}$ / $-2^{nd}$ & 
        \cite{google} &
        Skin-10 / Skin-100 \\ \hline
        Type  & D &  D &  D  & C  & C / C  & C  & C / C & C & C / C \\
        \#Classes & 3 & - & 7  & 6  & 3 / 7                   & 44                     & 128 / 198   & 26    &10 / 100      \\
        \#Images & 200 & 23,801 & 10,015   & 366 & 90 / 706                 & 2,309                   & 5,619 / 6,584   & 17,777   & 10,218 / 19,807   \\ 
        Annotation? & Y &  Y & N & N  & N / N & N & N / N & N &Y / N \\ 
        Year  & 2013  & 2016 & 2018 & 1998  & 2012          & 2013                   &  2016     &2019      & 2019  \\ \hline 
    \end{tabular}
\end{table*}

Unlike dermoscopic images, which are hard to obtain because of high expense and inconvenience access \cite{tsang2010global}, clinical images can be captured by easily-accessed devices, like the smartphone. However, there are several difficulties in recognizing skin lesion from a clinical image. (a) the background of clinical images is more complex so that how to reduce background noise interference is an important issue to consider. (b) clinical images cover far more classes of diseases than dermoscopic images, because dermatoscope is designed primarily for skin cancers, like melanoma and basal cell carcinoma, etc. (c) clinical skin disease image classification can be considered as a task of fine-grained classification, which is more difficult than normal classification task (e.g. CIFAR10 \cite{krizhevsky2009learning} and ImageNet \cite{deng2009imagenet}). On the one hand, the lesions of one class of skin diseases may not only appear in different parts of the human body, but also the visual characteristics of these lesions vary greatly, which indicates the high intra-class variance. On the other hand, two lesions that look very similar may belong to different categories, which indicates the low inter-class variance. Despite the above difficulties, it would be helpful for doctors and patients if we can apply deep learning techniques on our datasets to solve those problems.

To verify whether deep neural network works well on the task of clinical skin disease classification, we first benchmark on Skin-10 and Skin-100 using four SOTA CNNs: ResNet50 \cite{he2016deep}, DenseNet121 \cite{huang2017densely}, Nasnetamobile \cite{zoph2018learning} and Pnasnet5large \cite{liu2018progressive}, and we implement ensemble methods based on above four CNN models. Besides, we perform two SOTA object detection models (RetinaNet \cite{lin2017focal} and Faster-RCNN \cite{ren2015faster}) on Skin-10 to detect possible lesion regions on the raw images and then classify the disease based on the region with the highest confidence score. The experimental result shows that the object detection based method can reduce the influence of the background and achieve higher accuracy than plain CNN models. 

In summary, this paper has the following contributions. (a) We build two versions of clinical skin disease datasets: Skin-10 and Skin-100. As far as we know, Skin-10 is the first clinical skin disease dataset providing bounding boxes and the scale of Skin-100 is larger than most existing clinical skin disease datasets. (b) We establish the baseline performance of four SOTA CNN models on our datasets. (c) We verify the effectiveness of the ensemble method which achieves the best accuracy on both datasets. (d) We propose an object detection based approach and evaluate its performance on Skin-10. 

The remainder of this paper is organized as follows. Section \ref{section:related_work} reviews the existing related skin disease datasets and methods of recognizing skin diseases. In Section \ref{section:dataset}, the details of Skin-10 and Skin-100 are presented. Section \ref{section:method} describes the details of CNN-based, ensemble, and object detection based methods, and analyzes the experimental results. Finally, Section \ref{section:conclusion} offers concluding remarks and future works.

\section{Related Work}
\label{section:related_work}

\subsection{Skin Disease Datasets}


There are two types of skin disease images: dermoscopic and clinical. The former is obtained by a dermatoscope, which requires a high-quality magnifying lens and a powerful lighting system. Hence, dermoscopic images have a simpler background, which means that the distribution, number, size, shape, and color of skin lesions are much clearer compared with clinical images. Table \ref{table:dataset_summary} summarizes some existing skin disease datasets. PH2 dataset \cite{mendoncca2013ph} contains 200 dermoscopic images of melanocytic lesions, including 80 common nevi, 80 atypical nevi, and 40 melanomas. HAM10000 \cite{tschandl2018ham10000} collects dermoscopic images from different populations and consists of 10,015 images from 7 classes. ISIC Archive\footnote{https://isic-archive.com/} provides both dermoscopic and clinical images, which are categorized into many classes based on different attributes, such as diagnostic attributes, clinical attributes, etc. ISIC has 23,801 dermoscopic images but only has 100 clinical skin images. The dataset \cite{data1998} is from the UCI machine repository and contains six classes of skin diseases. In \cite{razeghi2012skin}, there are two datasets proposed: 90 dermatological images covering 3 skin disease classes and 706 images covering 7 classes, respectively. The dataset in \cite{razeghi2013interactive} has 44 classes containing 2,309 images. Another two datasets are proposed in \cite{sun2016benchmark}: SD-128, which has 128 classes and 5,619 images, and SD-198, which has 198 classes and 6,584 images. They achieve the classification accuracy of 52.15\% on SD-128 and 50.27\% on SD-198. A recent work proposed by Google \cite{google} develops a deep learning system for diagnosing 26 classes of clinical skin diseases using 17,777 cases. Each case consists of one or several clinical images and metadata, which includes patient demographic information and medical history.

\subsection{Skin Disease Classification}

The methods of skin disease image classification are two-folds. The first relies on hand-crafted features, such as the texture (SIFT, HOG), color (ColorSIFT, ColorHistogram), and edge (Gabor, Sobel) \cite{sun2016benchmark}.  The models, like support vector machines (SVMs), k-nearest neighbors (KNN) and decision trees, are then trained based on above features. The second is deep learning techniques. In \cite{sun2016benchmark}, a pretrained CNN model is used to extract deep features and the result shows that CNN performs better than hand-crafted based methods when the background is complex. In \cite{gessert2018skin}, the authors use a large ensemble of SOTA CNN models and place second at the ISIC 2018 challenge for skin lesion diagnosis. Verma and Pal et al. \cite{ensemble2019} also propose an ensemble method that combines five different data mining techniques. Zhang's work \cite{zhang2019attention} proposes an attention residual learning CNN model to effectively locate the skin lesion of dermoscopic images. Currently, the models generated by the neural architecture search (NAS) technique have been demonstrated to achieve comparable results to the human-designed models in many tasks \cite{automl,elsken2018neural}. In this paper, we compare NAS-designed and human-designed CNN models and perform the ensemble method based on these two types of models. We also evaluate the performance of two SOTA object detection models, which are expected to locate the skin lesion area and reduce the influence of background.

\section{Dataset}
\label{section:dataset}


%
%
%


\begin{figure}[!htb]
    \centering
    \includegraphics[width=0.48\textwidth]{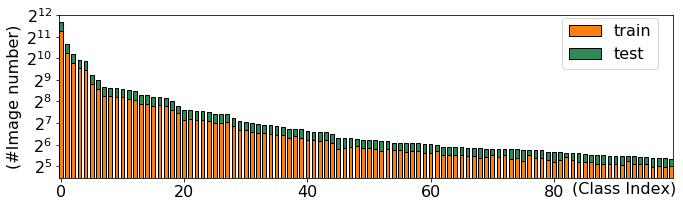}
        \caption{Statistics of Skin-100. A base-2 log scale is used for the Y-axis. For each class, the number of images in the training set is about 3 times the testing set.}
    \label{fig:data_distribution}
    
\end{figure}

\begin{table}[]
    \centering
    
    \caption{The statistics of Skin10. For notational convenience, we will use the index value to indicate the corresponding disease category in the following content unless otherwise stated.}
    \label{table:skin10_stat}
    \begin{tabular}{|c|c|c|c|}
        \hline
        Index & Class name & \#Training set & \#Testing set\\\hline
        0 & Acne Vulgaris  & 1598 & 399 \\\hline
        1 & Actinic Keratosis  & 932  & 239 \\\hline
        2 & Atopic Dermatitis Eczema  & 590  & 145 \\\hline
        3 & Basal Cell Carcinoma  & 3249  & 826 \\\hline
        4 & Compound Nevus  &  513 & 127 \\\hline
        5 & Onychomycosis  & 394  & 98 \\\hline
        6 & Rosacea  & 976  & 242 \\\hline
        7 & Seborrheic Keratosis  & 1180  & 291 \\\hline
        8 & Stasis Ulcer  &  407 & 100 \\\hline
        9 & Tinea Corporis  & 379  & 87 \\\hline
    \end{tabular}
\end{table}

    
    

Our datasets are built by scraping images from the Internet. We build Skin-10 by selecting 10 classes from the most common skin diseases \cite{hay2014global}. As a result, Skin-10 has 10,218 images and the statistics of Skin-10 is presented in Table \ref{table:skin10_stat}. Additionally, we use a graphical image annotation tool (LabelImg \footnote{https://github.com/tzutalin/labelImg}) to mark skin lesions with bounding boxes for each image in Skin-10. We further build a larger dataset based on Skin-10, namely Skin-100, which has 19,807 images of 100 skin diseases, and each class has over 40 images. The data distribution of Skin-100 is long-tailed, as shown in Fig \ref{fig:data_distribution}. In both Skin-10 and Skin-100, the ratio of the training set to the testing set is set at 3:1.
 

In some images, the skin lesion is covered or cured. Hence, we perform data cleaning to remove a total of 290 noise images.  The experimental result in Section 4 shows that data cleaning improves the performance of models for all CNN models.

\textbf{Scale} To the best of our knowledge, the scale of Skin-100 is much larger than existing clinical skin disease datasets. As Table \ref{table:dataset_summary} shows, the number of clinical skin disease images in Skin-100 is almost 3 to 200 times than other datasets. Besides, Skin-10 is the first clinical skin disease dataset providing bounding boxes for skin lesion detection \cite{sun2016benchmark,razeghi2012skin,razeghi2013interactive}. 



%
\textbf{Diversity} Our datasets cover different ages, genders and lesion locations. Besides, the difference can be significant within the same class, e.g. the skin disease images from the same class may differ from skin colors and lesion shapes, whereas, the difference can also be subtle between different classes. In a word, our datasets are of high diversity so that it is worthy to evaluate whether deep leaning techniques are feasible in our datasets.

%

%

\section{Clinical Skin Disease Image Classification}
\label{section:method}


In order to establish the baseline performance of CNN models on our datasets and not lose generality, we select four representative models from two types of models. (a) the models designed by human experts. (b) the models generated by the NAS algorithm. We also verify the feasibility of the ensemble method using these four models. We further evaluate the effectiveness of two SOTA object detection models. The implementation details and results are described in the following content.

\subsection{CNN based Classification}

\begin{figure}
    \centering
    \includegraphics[width=0.48\textwidth]{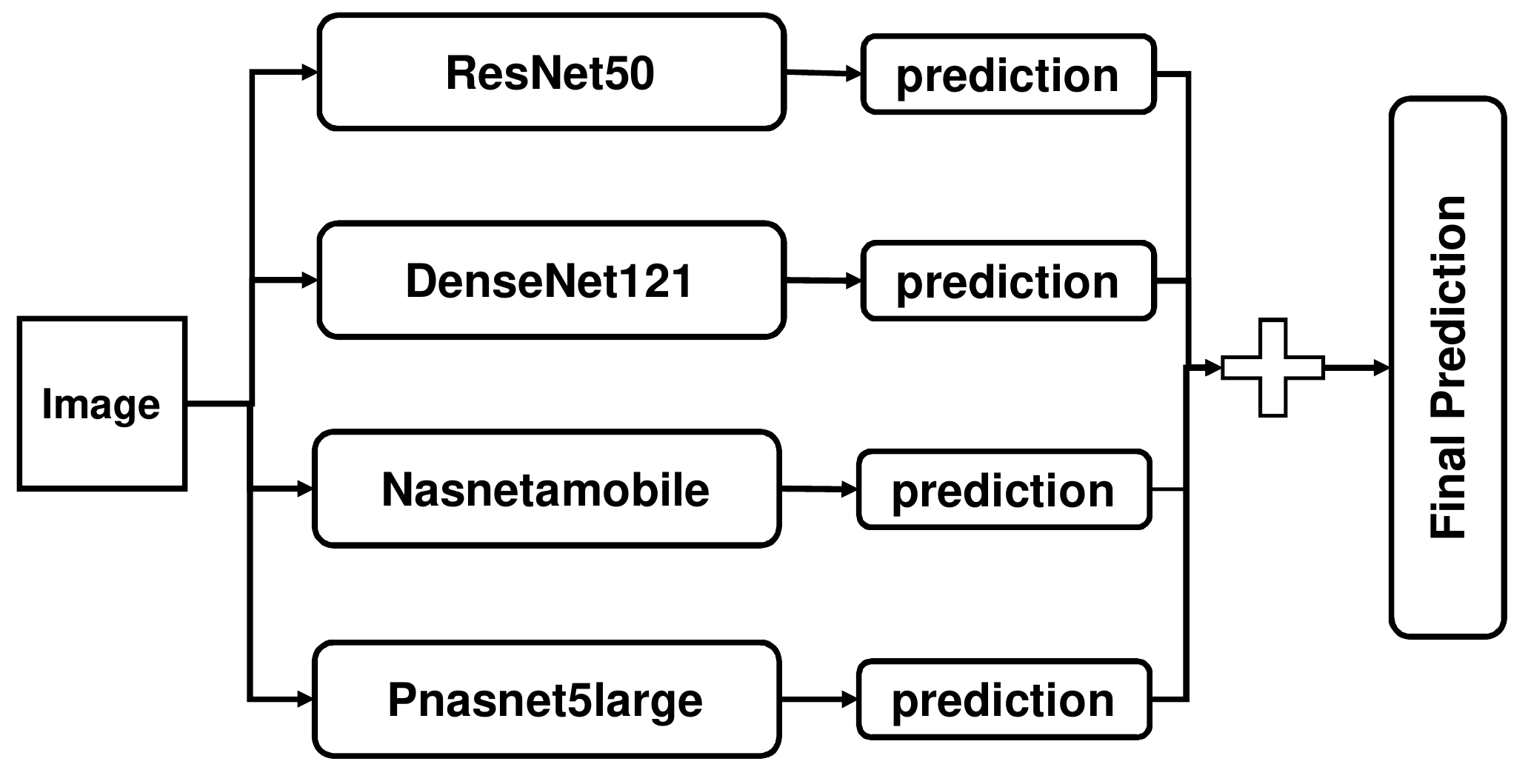}
    \caption{An overview of EnsembleNet.}
    \label{fig:ensemblenet}
\end{figure}

\begin{figure*}[!hbt]
    \centering
    \includegraphics[width=\textwidth]{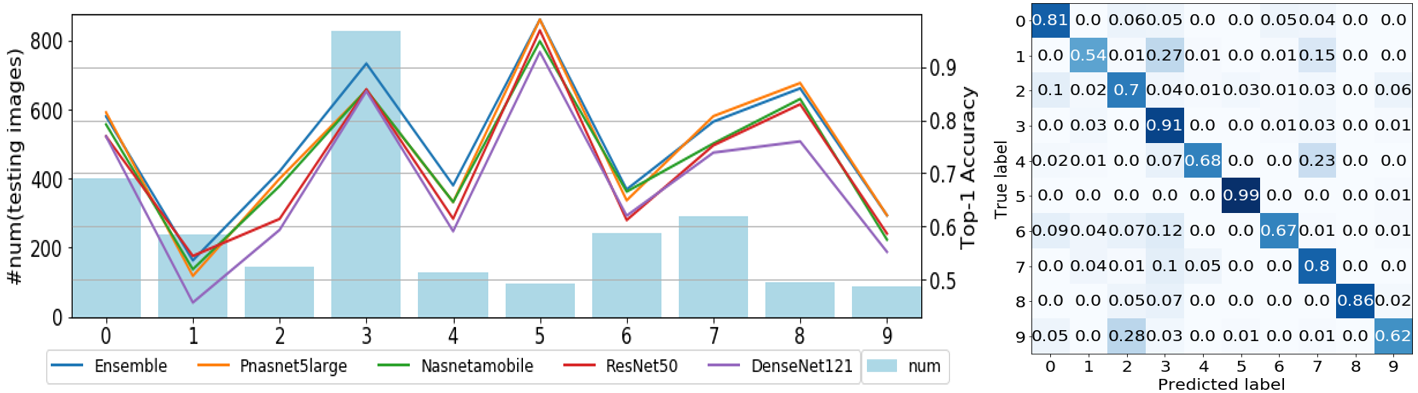}
    \caption{Left: The results of different models on Skin-10. Right: the confusion matrix of EnsembleNet. The number 0-9 in the x-axis indicates different class index.}
    \label{fig:skin10ensemble_analysis}
\end{figure*}

In this experiment, two types of SOTA CNN models are used as the baseline models: (a) ResNet50 and DenseNet121, which are designed by human experts. (b) Nasnetamobile and Pnasnet5large, which are generated automatically by the NAS technique. The pretrained models of (a) and (b) are obtained from torchvision\footnote{https://github.com/pytorch/vision/tree/master/torchvision/models} and pretrained-models.pytorch\footnote{https://github.com/Cadene/pretrained-models.pytorch}, respectively. After fine-tuning four base CNN models, we ensemble them into a strong classifier, namely EnsembleNet (shown in Fig \ref{fig:ensemblenet}), by summing the probability prediction of four base models:

\begin{equation}
    \text{EnsembleNet}(x)=\sum_k^K\text{BaseNet}_k(x)
\end{equation}
\noindent where $x$ indicates the input image and $K$ represents the number of base models. In this work, the best result is obtained by setting $K=4$, i.e. combining the prediction results of all base models.


\subsubsection{Settings} Before feeding the training set to the model, we implement a series of data augmentation, including resize, random crop, flips, rotation, and normalization, while for the testing set, we only perform resize and normalization. We use the stochastic gradient descent (SGD) optimizer with an initial learning rate of 0.01. The learning rate will be multiplied by a factor of 0.1 every 10 epochs. The batch size is 64 and the input image size is fixed to 224*224. Cross-entropy is used as the loss function. All baseline models are fine-tuned to converge.

\begin{table}
    \centering
    
    \caption{The classification accuracy of CNN models on our datasets. 
    }
    \label{table:benchmark}
    \begin{tabular}{|c|c|c|c|c|c|c|}
        \hline
        Dataset & \multicolumn{2}{c|}{Skin-10} &  \multicolumn{2}{c|}{Skin-100} &  \multicolumn{2}{c|}{Skin-100-Noise} \\\hline
        \begin{tabular}[c]{@{}c@{}}Top-k\\ Accuracy (\%)\end{tabular}  & Top-1  & Top-3  & Top-1  & Top-5 & Top-1  & Top-5    \\ \hline
        ResNet50   & 74.75 & 94.28 &   47.70   & 74.94 & 40.36 & 68.18        \\
        DenseNet121  & 72.94  & 92.68 & 48.43   & 75.37 & 45.08 &  70.76 \\ 
        Nasenetamobile  & 75.8  & 94.01 &  46.69 &  73.57 &  45.53 &  73.29 \\ 
        Pnasnet5large & 76.94 & 94.36 & 48.24 & 73.66 & 47.24 & 73.32 \\
        EnsembleNet & \textbf{79.01} & \textbf{95.34} & \textbf{53.54} & \textbf{79.18} & \textbf{51.96} & \textbf{78.51} \\
        \hline
    \end{tabular}
\end{table}

    

\subsubsection{Results and Analysis}

Based on the observation that the distribution of our datasets is imbalanced, we use top-k weighted accuracy as the metric. Let $\hat{y}_i$ be the output of the CNN models when the input is $x_i$ and $T^k_i$ be the labels of the $k$ largest elements in $\hat{y}_i$. $y_i$ is the ground-truth label of the $i$-th sample and $w_i$ is the corresponding sample weight. The top-k weighted accuracy on the testing set is given as

\begin{equation}
    A_{\mathrm{top}-k}=\sum_{i=1}^{N} Z_{i}^{k}w_i
\label{eq:top_k}
\end{equation}

\noindent where

\begin{equation}
\begin{aligned} w_{i} &=\frac{\sum_{j} 1\left(y_{j}=y_{i}\right)}{N} \\ Z_{i}^{k} &=\left\{\begin{array}{ll}{1,} & {y_{i} \in \mathbf{T}_{i}^{k}} \\ {0,} & {\text { otherwise }}\end{array}\right.\end{aligned} \notag
\label{eq:w_z}
\end{equation}

\noindent and $N$ is the number of images in the test set.

Table \ref{table:benchmark} presents the results of the baseline models on Skin-10, Skin-100 and Skin-100-Noise. Skin-100-Noise represents the dataset in which there exist many noise images in the training set, while Skin-100 is the opposite. Additionally, we make sure that both Skin-100 and Skin-100-Noise share the same testing set. In this way, we can fairly see from Table \ref{table:benchmark} that removing the noise images contributes to the improvement of classification accuracy for all CNN models, especially for ResNet50, as its top-1 accuracy is improved by 7.34\%. On the other hand, we can also find that, after data cleaning, the performance of models designed by the NAS technique is not improved as much as the human-designed models, which means that these models are more robust and able to extract more discriminative features. 

We further employ an ensemble method and we evaluate all possible combinations of base models. The experimental results show that the ensemble of four base models (EnsembleNet) achieves the highest top-1 accuracy (79.01\%), and the ensemble of DenseNet121, Nasnetamobile, and Pnasnet5large obtains the best top-3 accuracy (95.42\%). What's more, all ensemble models outperform any single base model.

\begin{table*}[!hbt]
    \centering
    
    \caption{The results of different models on Skin-10. The index 0 to 9 indicate the classification accuracy of different classes. The last column is the overall weighted accuracy on Skin-10.}
    \begin{tabular}{|c|c|c|c|c|c|c|c|c|c|c|c|}
        \hline
    \multicolumn{1}{|c|}{\multirow{2}{*}{Model}} & \multicolumn{11}{c|}{Top-1 Accuracy(\%)} \\ \cline{2-12} 
\multicolumn{1}{|c|}{} & \multicolumn{1}{c|}{0} & \multicolumn{1}{c|}{1} & \multicolumn{1}{c|}{2} & \multicolumn{1}{c|}{3} & \multicolumn{1}{c|}{4} & \multicolumn{1}{c|}{5} & \multicolumn{1}{c|}{6} & \multicolumn{1}{c|}{7} & \multicolumn{1}{c|}{8} & \multicolumn{1}{c|}{9} & Overall \\ \hline


ResNet50  & 76.94 & 54.39 & 61.38 & 85.84 & 61.42 & 96.94 & 61.16 & 75.26 & 83.00 & 58.62 & 74.75 \\\hline
DenseNet121  & 76.94  & 45.61 & 59.31 & 85.47 &  59.06& 92.86 & 61.98 & 73.88 & 76.00 &55.17  &72.94   \\ \hline
Nasnetamobile & 79.20  & 51.88 & 67.59 & 85.71 &  64.57 & 94.90 & 66.53 & 75.60 & 84.00 & 57.47  & 75.80   \\ \hline
Pnasnetamobile & \textbf{81.45} & 50.63 & 68.97 & 85.59 & 64.57 & \textbf{98.98} & 64.88 & 80.76 & \textbf{87.00} & 62.07 & 76.94 \\\hline
Faster-RCNN & 79.70 & 60.67 & 47.59 & 87.17 & \textbf{67.72} & \textbf{98.98} & \textbf{77.69} & 78.69 & 76.00 & 63.22 & 77.64 \\ \hline
RetinaNet  & 80.95 & \textbf{61.09} & \textbf{68.97} & \textbf{88.01} & 55.91 & 96.94 & 69.10 & \textbf{83.51} & 71.0 & \textbf{65.52} & \textbf{78.31} \\\hline
    \end{tabular}
    \label{table:obj_det}
\end{table*}

Although EnsembleNet can improve classification accuracy by integrating base models, we can see from Fig \ref{fig:skin10ensemble_analysis} (left) that EnsembleNet and base CNN models have similar performance on Skin-10, e.g. they all achieve high score on class 3, 5 and 8, but perform not well on class 1, 2, 4, and 9. Fig \ref{fig:skin10ensemble_analysis} (right) plots the confusion matrix of EnsembleNet, from which we can see that the classes that confuses EnsembleNet can be divided into two groups: (a) class 0, 2 and 9. (b) class 1, 3, 4, 6 and 7. The reason why the accuracy of class 0 and 3 is relatively higher probably is that they have much more images. But why is the accuracy of class 5 and 8 is high even though the number of images in both classes is not as many as class 0 and 3? Based on the observation of the images of different classes, we find that most of the lesion locations of class 5 and 8 are nails and legs, respectively, while the lesions of other classes are distributed in similar regions (e.g. faces, chest or unknown area) and their visual features also look similar. Hence, we can conclude that for class 5 and 8, most of their lesions are distributed in the specific regions, so the CNN models can correctly classify them even if it does not correctly locate the lesion regions. However, for some classes with similar lesion locations, the CNN models may misclassify the images due to focusing on the wrong region or failure to extract effective and discriminative features from the lesion area.

\subsection{Object Detection based Classification}

\begin{figure}
    \centering
    \includegraphics[width=0.48\textwidth]{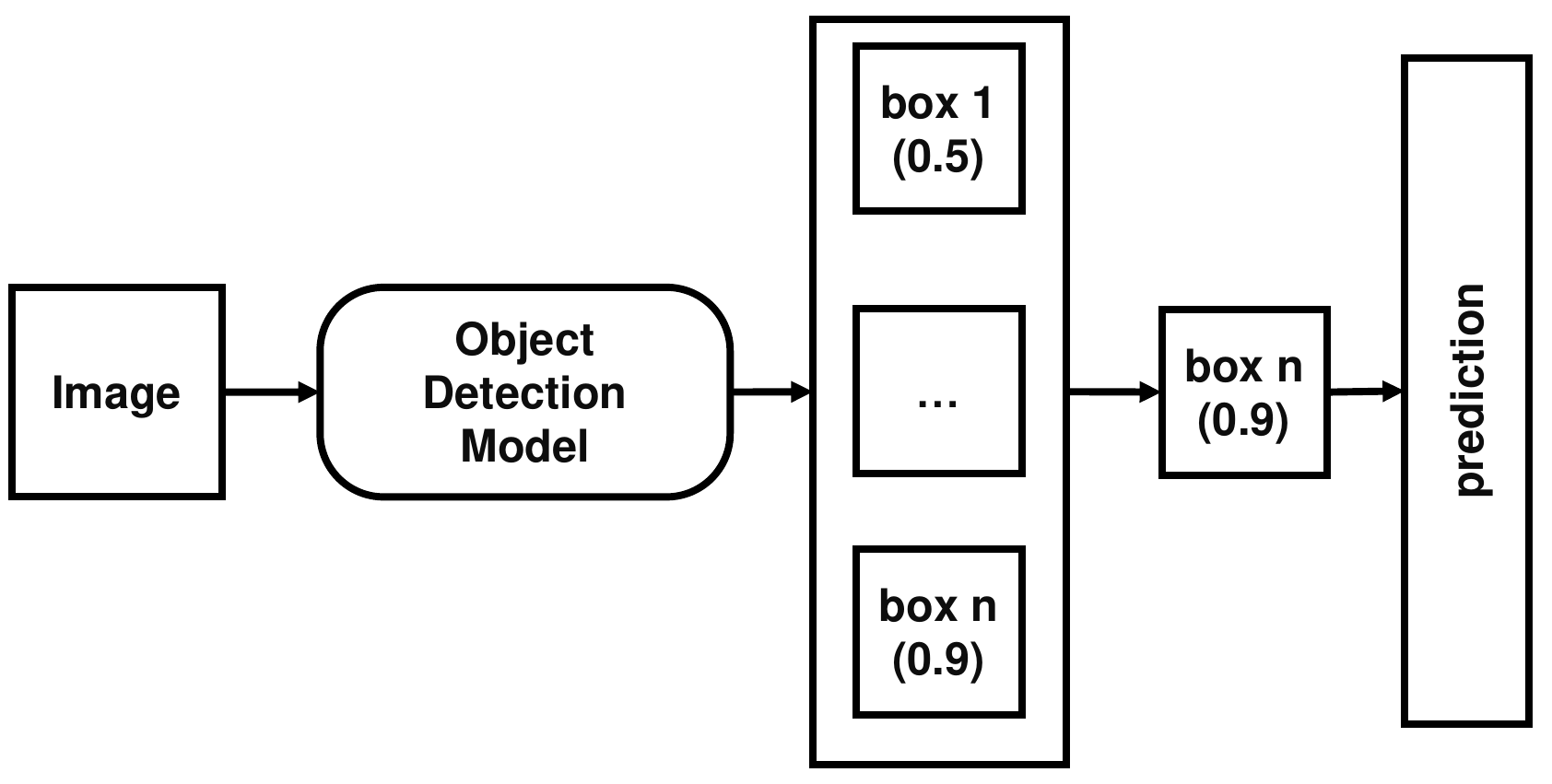}
    \caption{An overview of object detection based method. The final prediction is made based on the box with the highest confidence score.}
    \label{fig:obnet}
\end{figure}

To alleviate the problem of the plain CNN models, we evaluate the performance of two SOTA object detection models (RetinaNet and Faster-RCNN), respectively, which are expected to learn to detect and classify based on the skin lesion region. The process is shown in Fig \ref{fig:obnet}.

\subsubsection{Settings} The implementation of both RetinaNet and Faster-RCNN are publicly available in mmdetction\footnote{https://github.com/open-mmlab/mmdetection}. The backbone of both models is ResNeXt101\_64x4d. The input images are zoomed to 400$\times$400. The model is optimized by SGD with an initial learning rate of 0.001, and the learning rate is dropped by a factor of 10 every 10 epochs. Besides, for RetinaNet, the focal loss \cite{lin2017focal} is used to replace the commonly used cross-entropy loss, and we set $\alpha=0.25, \gamma=2$.

\subsubsection{Results and Analysis} Mean Average Precision (mAP) is one of the most commonly used metric for the object detection task. However, for Skin-10, each image corresponds to only one class of skin diseases, and we care more about whether the image is classified correctly. Therefore, we take the prediction of the box, which has the highest confidence, as the prediction of the image.

Let $\hat{y}_i$, $y_i$ be the prediction, ground truth for the $i$-th image, respectively. $w_i$ is the sample weight for $i$-th image, as defined in Equation \ref{eq:w_z}. $N$ and $C$ indicate the number of testing images and classes, respectively. Then the top-1 classification accuracy is defined as follows:

\begin{equation}
\begin{aligned} 
\text {Accuracy} &=\frac{\sum_{i}^{N} \mathbb{1}\left(\hat{y}_{i}, y_{i}\right)}{N}, \mathbb{1}(p, q)=\left\{\begin{array}{ll}{1} & {\text { if } p=q} \\ {0} & {\text { if } p \neq q}\end{array}\right.\\ 
\text { s.t. } \hat{y}_{i} &=\underset{n}{\arg \max }\,\, b_{i}^{m n}, m \in\left[1, M_{i}\right], n \in\left[1, C\right] 
\end{aligned}
\end{equation}

\noindent where $M^i$ represents the number of predicted boxes in the $i$-th image. $b_{i}^{m n}$ indicates the probability of being predicted as the $n$-th class for the $m$-th box in the $i$-th image.

    


It can be seen from the result in Table \ref{table:obj_det} that the classification accuracy of most classes is improved by detecting skin lesions. However, we can also find that the classification accuracy of some classes (e.g. class 8) decreases significantly for both RetinaNet and Faster-RCNN, instead. There are several possible explanations for this result. First, the prediction is only based on the box with the highest confidence. In other words, it loses the global features that are important for classification, which further confirms the conclusion of the previous experiment that the reason why base CNN models can achieve higher classification accuracy in some classes (e.g. class 8) is because lesion locations of these classes are unique and not easy to be confused by other classes. Second, although both object detection models can detect lesions well in most cases, it may detect the wrong area or even fail to detect any targets when the skin lesion area is not clear or affected by background noise, thus leading to failure of prediction.

\section{Conclusions and Future Work}
\label{section:conclusion}


In this paper, we investigated how deep learning techniques could help image-based skin disease diagnosis. We developed two versions of clinical skin disease datasets from Internet images: Skin-10, which consists of 10,218 images of 10 common skin disease classes with bounding boxes surrounding the lesion, and Skin-100, which contains 19,807 images of 100 skin disease classes. We found that data cleaning is very important as it can help improve the top-1 accuracy by 4\% on average in our experiments. We also found that the ensemble method makes more efficient use of the dataset and outperforms any single CNN model on both Skin-10 and Skin-100. We further evaluated two SOTA object detection models that are used to reduce the influence of image background. Our results showed that the object detection models outperform the classification based solutions. 

Although it is demonstrated that deep learning techniques can achieve satisfactory performance in skin disease image classification, there still exists room for further improvement. Based on the analysis of previous experimental results, we summarize the following directions worth studying in the future. First, the models generated by autoML technique may extract more discriminative features and are more robust to noisy data than human-designed models, therefore we can try to explore autoML technique to generate a task-specific architecture based on clinical skin disease datasets. Second, although the ensemble method is easy to implement and effective, it does not improve the accuracy of some hard-to-classify classes. Thus, how to improve the accuracy of these classes is still a challenging task. Third, object detection based approach can reduce the influence of background by detecting the local region of skin lesions, but it loses the global features. Hence, how to combine global and local features would be another promising work. At last, our dataset is built on Internet images and cannot be open to the public due to copyright issue. We hope the healthcare community can join hands to develop an open and large skin disease dataset.






\bibliographystyle{IEEEtran}
\bibliography{IEEEabrv,conference_101719}

\end{document}